\documentclass[10pt]{article}                           
\usepackage{graphics} 

\def \Oeuvres{O$\!$euvres}
\def \ie {i.e.~} 
\def\LHS{l.h.s.~}
\def\RHS{r.h.s.~}
\def \ccomma{\raise 2pt\hbox{,}} 

\def \Log {\mathop{\rm Log}\nolimits}
\def \degree{\mathop{\rm degree}\nolimits}

\def\CRAS{C.~R.~Acad.~Sc.~Paris}

\def\SAM{Stud.~Appl.~Math.~}
\def\Pn     {{\rm Pn}}
\def\PI     {{\rm P1}}
\def\PII    {{\rm P2}}

\def\PV     {{\rm P5}}
\def\PVI    {{\rm P6}}

\def\Alpha{A}

\def\Beta {B}
\def\abcd{\alpha,\beta,\gamma,\delta}
\def\ABCD{\Alpha,\Beta,\Gamma,\Delta}

\def\veca{\mbox{\boldmath{$\alpha$}}}
\def\vecA{\mbox{\boldmath{$\Alpha$}}}
\def\vect{\mbox{\boldmath{$\theta$}}}
\def\vecT{\mbox{\boldmath{$\Theta$}}}

\def\BiT{birational transformation}

\def \TPVI  {{\rm T}_{\rm 6}}     
\def \Hbadc {{\rm H}_{\rm badc}}  
\def \Hdcba {{\rm H}_{\rm dcba}}  
\def \Hcdab {{\rm H}_{\rm cdab}}  

\def\Mugan{Mu\u gan}

\def\vplus {     \bar{v}}
\def\vminus{\underbar{$v$}}

\begin{document}

\title{
New contiguity relation of the sixth Painlev\'e equation from a truncation
\footnote{To appear in Physica D. Preprint S2001/009.
Corresponding author RC, 
e-mail Conte@drecam.saclay.cea.fr,
phone +33--1--69087349,
fax   +33--1--69088786.
}
}

\author{Robert Conte\dag\ and Micheline Musette\ddag
\\ \dag Service de physique de l'\'etat condens\'e, CEA--Saclay
\\ F--91191 Gif-sur-Yvette Cedex, France
\\ E-mail:  Conte@drecam.saclay.cea.fr
\\ \ddag Dienst Theoretische Natuurkunde, Vrije Universiteit Brussel
\\ Pleinlaan 2, B--1050 Brussels, Belgium
\\ E-mail:  MMusette@vub.ac.be
}

\maketitle

\medskip

\begin{abstract} 
For the master Painlev\'e equation $\PVI(u)$,
we define a consistent method,
adapted from the Weiss truncation for partial differential equations,
which allows us to obtain the first degree \BiT\ of Okamoto.
Two new features are implemented to achieve this result.
The first one is the homography between the derivative of the solution $u$
and a Riccati pseudopotential.
The second one is an improvement of a conjecture by Fokas and Ablowitz 
on the structure of this \BiT.
We then build the contiguity relation of $\PVI$,
which yields one new second order nonautonomous discrete equation.
\end{abstract}


\medskip


\hfill 

\tableofcontents
\vfill \eject

\baselineskip=14truept


\section{Introduction}

Second order first degree algebraic ordinary differential equations (ODEs)
define six and only six functions \cite{GambierThese},
which satisfy neither a first order ODE nor a linear ODE.
Since the sixth of these \textit{Painlev\'e equations} $\Pn$ can generate
the five others by a confluence process \cite{PaiCRAS1906},
we will mainly concentrate on it
\begin{eqnarray*}
\label{eqP6}
\PVI\ : \
u''
&=&
\frac{1}{2} \left[\frac{1}{u} + \frac{1}{u-1} + \frac{1}{u-x} \right] {u'}^2
- \left[\frac{1}{x} + \frac{1}{x-1} + \frac{1}{u-x} \right] u'
\\
& &
+ \frac{u (u-1) (u-x)}{x^2 (x-1)^2} 
  \left[\alpha + \beta \frac{x}{u^2} + \gamma \frac{x-1}{(u-1)^2} 
        + \delta \frac{x (x-1)}{(u-x)^2} \right]\cdot
\end{eqnarray*}
It depends on four arbitrary complex constants $\abcd$.

The purpose of this article is twofold :
(i)
we first present a new, direct method,
only based on the singularity structure of this ODE,
in order to derive a first degree \textit{birational transformation}
for $\PVI$;
(ii)
we then establish for $\PVI$ the contiguity relation,
interpreted as a discrete Painlev\'e equation.

A birational transformation is defined by
\begin{eqnarray} 
& &
u = r(U',U,X),\
U = R(u',u,x),
\label{eqbira}
\end{eqnarray}
with $r$ and $R$ rational functions,
and it maps an equation
\begin{eqnarray} 
& &
E(u) \equiv \Pn(u,x,\veca)=0,\
\veca=(\abcd),\
\label{eqPnu}
\end{eqnarray}
into the same equation with different parameters
\begin{eqnarray} 
& &
E(U) \equiv \Pn(U,X,\vecA)=0,\
\vecA=(\ABCD),
\label{eqPnU}
\end{eqnarray}
with some homography (usually the identity) between $x$ and $X$.
The parameters $(\veca,\vecA)$ must obey as many algebraic relations
as elements in $\veca$.
The \textit{degree} of a \BiT\ is by definition the highest degree in $U'$
or $u'$ of the numerator and the denominator of (\ref{eqbira}).

There are at least three uses of a \BiT.
The first one is an equivalence relation:
two solutions exchanged by the transformation are considered as identical.
The second use is to generate from a known solution a countable number 
of new solutions which may be of physical interest \cite{Maz2001a}.
The third point of view,
which we develop here, 
is the construction of a contiguity relation
in order to define a nonautonomous discrete Painlev\'e equation.

The interest of being able to handle $\PVI$ is to avoid the tedious
consideration of the other daughter Painlev\'e equations.

Indeed, all the similar results for these equations can be easily
obtained by action of the confluence,
among them all the first degree {\BiT}s of all $\Pn$ equations
\cite{CM2001c}.

There currently exist two {\BiT}s of $\PVI$.
Garnier \cite{Garnier1943a,Garnier1951}
was the first to find a birational transformation 
of $\PVI$,
to establish a theorem of Schwarz on the problem of Plateau.
This transformation, which has second degree,
was later rediscovered by several authors 
\cite{FY,Okamoto1987I,MS1995b}.
There also exists a first degree \BiT,
found by Okamoto \cite{Okamoto1987I}
while studying the Weyl group which preserves the Hamiltonian of $\PVI$,
and rediscovered recently \cite{NJH}
in a different context.
This transformation reads
\begin{eqnarray}
\TPVI : \
\frac{N}{u-U}
& = &
  \frac{x (x-1) U'}{U (U-1)(U-x)}
 +\frac{\Theta_0}{U}+\frac{\Theta_1}{U-1}+\frac{\Theta_x-1}{U-x}
\label{eqTP6uvecTUnsigned}
\\
& = &
  \frac{x(x-1)u'}{u(u-1)(u-x)}
 +\frac{\theta_0}{u}+\frac{\theta_1}{u-1}+\frac{\theta_x-1}{u-x}\ccomma
\label{eqTP6uvectUnsigned}
\\
\theta_j
& = &
\Theta_j - \frac{1}{2} \left(\sum \Theta_k\right) + \frac{1}{2}
\ccomma\
j,k=\infty,0,1,x,
\label{eqT6}
\\
\Theta_j
& = &
\theta_j - \frac{1}{2} \left(\sum \theta_k\right) + \frac{1}{2}
\cdot
\label{eqT9}
\end{eqnarray}
The transformation is clearly birational since the \LHS\ is homographic
in both $u$ and $U$.
In the above, the monodromy exponents
$\vect=(\theta_\infty,\theta_0,\theta_1,\theta_x)$
are defined as
\begin{eqnarray} 
& &
\theta_\infty^2= 2 \alpha,\
\theta_0^2     =-2 \beta,\
\theta_1^2     = 2 \gamma,\
\theta_x^2     =1 - 2 \delta,
\end{eqnarray}
and similarly for their uppercase counterparts,
while the odd-parity constant $N$ takes the equivalent expressions
\begin{eqnarray}
N
& = &
\sum (\theta_k^2 - \Theta_k^2)
\label{eqP6N}
\\
& = &
  1 - \sum     \Theta_k
=
 -1 + \sum     \theta_k
\label{eqNvect}
\\
& = &
  2 (    \theta_j -     \Theta_j),\ j=\infty,0,1,x.
\label{eqNvecT}
\end{eqnarray}

The transformation of Garnier is an integer power \cite{C2001TSP6}   
of this first degree transformation.

To achieve our goal (rely only on the singularity structure to find a \BiT),
we need to improve the \textit{singular manifold method}
so that it succeeds to obtain a \BiT\ for $\PVI$.
Originally introduced for partial differential equations (PDEs)
by Weiss, Tabor and Carnevale \cite{WTC},
the singular manifold method is a powerful tool
for deriving B\"acklund transformations,
by considering only the singularity structure of the solutions.
Its current achievements are detailed in summer school proceedings,
see Refs.~\cite{Cargese96Musette,CetraroConte}.
An extension to ODEs has been proposed \cite{CJP,GJP1999a,GJP2001N} 
to derive a \BiT\ for the Painlev\'e equations,
but its application to the master equation $\PVI$ is still an open problem.
We solve it here by implementing an essential piece of information,
which has up to now been overlooked.

The paper is organized as follows.
In Section \ref{sectionHomography},
we exploit the information that there always exists a homography between 
the derivative 
of the solution 
of the considered Painlev\'e equation $\Pn$
and the Riccati pseudopotential $Z$
introduced in the ``truncation'' assumption.
This reduces the problem to finding two functions of two variables
instead of two functions of three variables.

In Section \ref{sectionOneFamilyTruncation},
we implement this homography in the definition of a truncation,
which allows us
to overcome the major difficulty, coming, in the case of $\PVI$,
from a residual coefficient which remains undetermined
(technically coming from the value $1$ of the Fuchs index).
This difficulty is the main reason why $\PVI$ has never been handled before
by methods based on the singularity structure.
We also improve a previous conjecture \cite{FA1982}
on the necessary form of the \BiT.
The truncation then becomes easy to solve and admits,
up to the four homographies on $(U,x)$ which conserve $x$,
a unique solution, which is the transformation $\TPVI$ of Okamoto.

In Section \ref{sectionElementarySTPVI},
we give the various representations of this transformation.

In Section \ref{sectionDiscrete},
we solve the recurrence relation between the monodromy exponents of $\PVI$,
and we build the contiguity relation,
which defines a unique nonautonomous second order difference equation.

Finally, in Section \ref{sectionOnST},
we propose a weaker definition of a \textit{Schlesinger transformation}
so as to identify this notion with that of birational transformation.

\vfill \eject

\section{The fundamental homography}
\label{sectionHomography}

{}From the result of Richard Fuchs \cite{FuchsP6},
the $\PVI$ equation is obtained from the complete integrability
of a linear differential system of order two.
Therefore the pseudopotential of the singular manifold method
has only one component $Z$ \cite{MC1991}, 
which can be chosen so as to satisfy some Riccati ODE.

Each $\Pn$ equation which admits a \BiT\ has one or several
(four for $\PVI$)
couples of families of movable simple poles with opposite residues $\pm u_0$,
therefore both the one-family truncation and the two-family truncation
\cite{MC1994,Cargese96Musette,CetraroConte}
are applicable.
In the present paper,
we consider only the one-family truncation,
whose assumption is
\begin{eqnarray} 
& &
u=u_0 Z^{-1} +U,\ u_0 \not=0,\
x=X,
\label{eqDTOne}
\\
& &
Z'=1 + z_1 Z + z_2 Z^2,\ z_2 \not=0,
\label{eqRiccatiOneFamily}
\end{eqnarray}
in which $u$ and $U$ satisfy (\ref{eqPnu}) and (\ref{eqPnU}),
$(Z,z_1,z_2)$ 
are rational functions of $(x,U,U')$ to be determined.
After this is done, the relation (\ref{eqDTOne}) 
represents half of the birational transformation.

Besides the equation (\ref{eqRiccatiOneFamily}),
there exists a second Riccati equation in the present problem,
this is the Painlev\'e equation (\ref{eqPnU}) itself.
Indeed, any $N$-th order, first degree ODE with the Painlev\'e property
is necessarily \cite[pp.~396--409]{PaiLecons}
a Riccati equation for $U^{(N-1)}$,
with coefficients depending on $x$ and the lower derivatives of $U$,
in our case
\begin{eqnarray}
& &
U''=A_2(U,x) U'^2 + A_1(U,x) U' + A_0(U,x).
\label{eqRiccatiUprime}
\end{eqnarray}

Since the group of invariance of a Riccati equation is the homographic group,
the variables $U'$ and $Z$ are linked by a homography,
the three coefficients $g_j$ of which are rational in $(U,x)$.
Let us define it as
\begin{eqnarray}
& &
 (U' + g_2) (Z^{-1} - g_1) - g_0=0,\ g_0 \not=0.
\label{eqHomographyRUprime}
\end{eqnarray}

This allows us to obtain the two coefficients
 $z_j$ of the Riccati pseudopotential equation (\ref{eqRiccatiOneFamily})
as explicit expressions 
of $(g_j, \partial_U g_j, \partial_x g_j, A_2,A_1,A_0,U')$.
Indeed,
eliminating $U'$ between (\ref{eqRiccatiUprime}) and
(\ref{eqHomographyRUprime}) defines a first order ODE for $Z$,
whose identification with (\ref{eqRiccatiOneFamily})
\textit{modulo} (\ref{eqHomographyRUprime})
provides three relations.

For the one-family truncation, these are
\begin{eqnarray} 
& &
g_0= g_2^2 A_2 - g_2 A_1 + A_0 + \partial_x g_2 - g_2 \partial_U g_2,
\label{eqg0Apriori}
\\
& &
z_1=
   A_1 - 2 g_1 + \partial_U g_2 - \partial_x \Log g_0 
+ \left(2 A_2 - \partial_U \Log g_0\right) U',
\label{eqz1Apriori}
\\
& &
z_2= -g_1 z_1 - g_1^2 - g_0 A_2 - \partial_x g_1 - (\partial_U g_1) U'.
\label{eqz2Apriori}
\end{eqnarray}

Therefore, the natural unknowns in the present problem are the two
coefficients $g_1,g_2$ of the homography,
which are functions of the two variables $(U,x)$,
and not the two functions $(z_1,z_2)$ of the three variables $(U',U,x)$.

\textit{Remark}.
One must also consider the case when the relation between $Z^{-1}$ and $U'$
is affine,
excluded in (\ref{eqHomographyRUprime}).
Assuming
\begin{eqnarray}
& &
G_1 (U' + G_2) - Z^{-1} =0,\ G_1 \not=0,
\label{eqLinearZUprime}
\end{eqnarray}
the equation analogous to (\ref{eqg0Apriori}) is now
\begin{eqnarray}
& &
\partial_U G_1 + G_1^2 + A_2 G_1=0,
\end{eqnarray}
which for $\PVI$ admits no solution $G_1$ rational in $U$.

\section{The truncation}
\label{sectionOneFamilyTruncation}

Just like the field $u$ is represented, see Eq.~(\ref{eqDTOne}),
by a Laurent series in $Z$ which terminates (``truncated series''),
the \LHS\ $E(u)$ of the $\Pn$ equation 
(supposed written as a polynomial, \ie\ without denominators involving $u$)
can also be written as a truncated series in $Z$.
This is achieved 
by the elimination of $u,Z',U'',U'$
between (\ref{eqPnu}), (\ref{eqPnU}), (\ref{eqDTOne}), 
(\ref{eqRiccatiOneFamily})
and (\ref{eqHomographyRUprime}),
followed by the elimination of $(g_0,z_1,z_2)$
from (\ref{eqg0Apriori})--(\ref{eqz2Apriori})
($q$ denotes the singularity order of $\Pn$ written as a 
differential polynomial in $u$, it is $-6$ for $\PVI$),
\begin{eqnarray}
& &
E(u) = \sum_{j=0}^{-q+2} E_j(U,x,u_0,g_1,g_2,\veca,\vecA) Z^{j+q-2}=0,
\label{eqST1LaurentE}
\\
& &
\forall j\ :\            E_j(U,x,u_0,g_1,g_2,\veca,\vecA)=0.
\label{eqST1Determining}
\end{eqnarray}
The nonlinear \textit{determining equations} $E_j=0$ are independent of $U'$,
and this is the main difference with previous work.
In particular,
in all cases successfully processed to date \cite{GJP1999a,GJP2001N},
both $g_1$ and $2 A_2 - \partial_U \Log g_0$ vanish,
which cancels the coefficient of $U'$ in $z_1$ and $z_2$.
In the case of $\PVI$, 
we are going to see that the truncation possesses a solution if and only if 
$(z_1,z_2)$ depends on $U'$.

Another difference is the greater number ($-q+3$ instead of $-q+1$)
of equations $E_j=0$,
which is due to the additional elimination of $U'$ with 
(\ref{eqHomographyRUprime}).

The $-q+3$ determining equations (\ref{eqST1Determining}) 
in the three unknown functions $u_0(x)$, $g_1(U,x)$, $g_2(U,x)$ 
(and the unknown scalars $\abcd$ in terms of $\ABCD$)
must be solved,
as usual, by increasing values of their index $j$.

This truncation for $\PVI$ possesses a very nice invariance,
which drastically helps the practical resolution.
For any solution $g_j(U,x)$, there exist three other solutions,
generated by the action on $g_j(U,x)$ of the 
four homographies of $U$ which conserve $x$ and $\PVI$, namely
the identity and
\begin{eqnarray}
& &
\Hbadc : \vect = 
 \pmatrix{ 0 & 1 & 0 & 0 \cr
           1 & 0 & 0 & 0 \cr
           0 & 0 & 0 & 1 \cr
           0 & 0 & 1 & 0 \cr} \vecT,\
x=X,\
u=\frac{x}{U}\ccomma
\label{eqHbadc}
\\
& &
\Hdcba : \vect = 
 \pmatrix{ 0 & 0 & 0 & 1 \cr
           0 & 0 & 1 & 0 \cr
           0 & 1 & 0 & 0 \cr
           1 & 0 & 0 & 0 \cr} \vecT,\
x=X,\
u-x=\frac{x(x-1)}{U-x}\ccomma
\label{eqHdcba}
\\
& &
\Hcdab : \vect = 
 \pmatrix{ 0 & 0 & 1 & 0 \cr
           0 & 0 & 0 & 1 \cr
           1 & 0 & 0 & 0 \cr
           0 & 1 & 0 & 0 \cr} \vecT,\
x=X,\
u-1=\frac{1-x}{U-1}\ccomma
\label{eqHcdab}
\\
& &
\Hbadc^2=\Hdcba^2=\Hcdab^2=1,\
\Hbadc   \Hdcba   \Hcdab = 1.
\end{eqnarray}

Let us now state a conjecture which will strongly restrict
the coefficient $g_2$.
{}From the expression of the direct half of the \BiT,
\begin{eqnarray}
& &
u=U + u_0 \left(g_1(U,x) + \frac{g_0(U,x)}{U'+g_2(U,x)}\right),
\label{eqSTDirecte}
\end{eqnarray}
the values of $U$ which correspond to $u=\infty$ 
are defined (apart from the poles of $g_1$ and $g_0$) by the ODE
\begin{eqnarray}
& &
U'+g_2(U,x)=0.
\label{eqRiccatiDenom}
\end{eqnarray}
The point $u=\infty$ has the peculiarity to be a singular point
of all the $\Pn$ equations and,
for all known {\BiT}s of $\Pn$
(see the book \cite{GLBook} for $\PII$ to $\PV$,
      Ref.~\cite[formula (2.8)]{Garnier1951}
  and Ref.~\cite[p.~356]{Okamoto1987I} for $\PVI$),
it happens that the ODE analogous to (\ref{eqRiccatiDenom}) is a Riccati ODE
(or a product of Riccati ODEs in \cite{Garnier1951,MS1995b}),
\ie\ the unique first order first degree ODE which has the 
Painlev\'e property.
In at least one other example of higher order \cite{Hone1998HH},
the birational transformation between two different 
ODEs having the Painlev\'e property
has a denominator which defines a $\PI$ equation.
Let us conjecture the generality of this property.
\medskip

\noindent \textbf{Conjecture}. 
\textit{
Given a birational transformation between two ODEs having
the Painlev\'e property,
for any singular point $u$ of the ODE,
the ODE for $U$ defined by the direct half of the \BiT\
has the Painlev\'e property.
}
\medskip

This is an improvement of a previous conjecture by Fokas and Ablowitz
\cite[formula (2.6)]{FA1982}
in two respects:
both fields $u$ and $U$ are required to satisfy the same ODE,
and no specific $U$-dependence is assumed for $g_0$ and $g_1$;
their conjecture 
(whatever be $u$, Eq.~(\ref{eqSTDirecte}) is a Riccati equation for $U$)
happens to be true for $\PII$--$\PV$ but not for $\PVI$.

In the case of $\PVI$,
our conjecture implies that the four expressions
\begin{eqnarray}
& &
g_2,\
\frac{g_0}{U+u_0 g_1},\
\frac{g_0}{U-1+u_0 g_1},\
\frac{g_0}{U-x+u_0 g_1},\
\end{eqnarray}
are second degree polynomials of $U$ with coefficients depending on $x$.
One notices immediately, from the last three fractions,
that there \textit{could} exist a particular solution
$g_1=0,g_0=g(x) U(U-1)(U-x)$.
We are going to see that this is indeed the case.

The practical resolution of the truncation for $\PVI$ is performed in Appendix 
\ref{appendixDetailedResultsOneFamilyTruncationPVI},
and the coefficients of the fundamental homography are found to be
\begin{eqnarray}
g_0
& = &
\frac{N U (U-1)(U-x)}{u_0 x(x-1)},\
g_1=0,\
g_2=
\frac{U(U-1)(U-x)}{x(x-1)}
\left(\frac{\Theta_0}{U}+\frac{\Theta_1}{U-1}+\frac{\Theta_x-1}{U-x}\right)
\ccomma\
\\
u_0
& = &
-\frac{x(x-1)}{\theta_\infty}\ccomma\
\theta_\infty = \frac{1}{2}
  \left( \Theta_\infty - \Theta_0 - \Theta_1 - \Theta_x + 1\right),
\label{eqg0g1g2}
\\
N
& = &
1-\Theta_\infty-\Theta_0-\Theta_1-\Theta_x,
\\
z_1 
& = &
 \frac{1}{x(x-1)} ((\Theta_1  +\Theta_x-1)  U
                  +(\Theta_x-1+\Theta_0  ) (U-1)
                  +(\Theta_0  +\Theta_1  ) (U-x))
\nonumber
\\
& &
+ \frac{1}{x}+ \frac{1}{x-1}\ccomma
\label{eqz1sol}
\\
z_2
& = &
   \frac{N \theta_\infty}{2(x(x-1))^2} ((U-1)(U-x)+U(U-x)+U(U-1)).
\label{eqz2sol}
\end{eqnarray}

Let us present this \BiT\ in more detail.

\section{The elementary \BiT\ of $\PVI$} 
\label{sectionElementarySTPVI}

The eight signs $(s_{\infty},s_0,s_1,s_x)$ and $(S_{\infty},S_0,S_1,S_x)$,
with $s_j^2=S_j^2=1$,
of the monodromy exponents remain arbitrary and independent.

\subsection{The affine representation} 

In the space of the monodromy exponents,
the direct \BiT\ and its inverse have an \textit{affine representation}
\begin{eqnarray}
& &
\vect = M_1 \vecT + M_0,\
\vect=\pmatrix{\theta_{\infty} \cr \theta_0 \cr \theta_1 \cr \theta_x},\
\vecT=\pmatrix{\Theta_{\infty} \cr \Theta_0 \cr \Theta_1 \cr \Theta_x},\
\label{eqSTM1M0}
\end{eqnarray}
in which $M_1$ and $M_0$ are matrices of rational numbers.
These are
\begin{eqnarray}
& &
\TPVI : 
\pmatrix{s_\infty \theta_\infty \cr
         s_0        \theta_0        \cr
         s_1        \theta_1        \cr
         s_x        \theta_x        \cr}
= \frac{1}{2}               \pmatrix{ 1 & -1 & -1 & -1 \cr
                                     -1 &  1 & -1 & -1 \cr
                                     -1 & -1 &  1 & -1 \cr
                                     -1 & -1 & -1 &  1 \cr}
\pmatrix{S_\infty \Theta_\infty \cr
         S_0        \Theta_0        \cr
         S_1        \Theta_1        \cr
         S_x        \Theta_x        \cr}
              + \frac{1}{2} \pmatrix{ 1 \cr 1 \cr 1 \cr 1 \cr}\ccomma
\\
& &
\TPVI^{-1}:
\pmatrix{S_\infty \Theta_\infty \cr
         S_0        \Theta_0        \cr
         S_1        \Theta_1        \cr
         S_x        \Theta_x        \cr}
= \frac{1}{2}               \pmatrix{ 1 & -1 & -1 & -1 \cr
                                     -1 &  1 & -1 & -1 \cr
                                     -1 & -1 &  1 & -1 \cr
                                     -1 & -1 & -1 &  1 \cr}
\pmatrix{s_\infty \theta_\infty \cr
         s_0        \theta_0        \cr
         s_1        \theta_1        \cr
         s_x        \theta_x        \cr}
              + \frac{1}{2} \pmatrix{ 1 \cr 1 \cr 1 \cr 1 \cr}\ccomma
\end{eqnarray}
and, depending on the signs,
one has $M_1^{n}=1$,
with $n=2,3,4$ or $6$.
The convention adopted for the signs is aimed at making (\ref{eqT6}) 
equal to its inverse when all signs are $+1$.

\subsection{The birational representation} 

Denoting $N$ the odd-parity constant
\begin{eqnarray}
N
& = &
\sum (\theta_k^2 - \Theta_k^2)
\\
& = &
  1 - \sum S_k \Theta_k
=
 -1 + \sum s_k \theta_k
\\
& = &
  2 (s_j \theta_j - S_j \Theta_j),\ j=\infty,0,1,x,
\end{eqnarray}
the four algebraic relations between $\abcd$ and $\ABCD$ are
\begin{eqnarray}
& &
\forall j=\infty,0,1,x\ : \
(\theta_j^2+ \Theta_j^2 - (N/2)^2)^2 - (2 \theta_j \Theta_j)^2=0,
\label{eqT6algebraic}
\end{eqnarray}
equivalent to the affine representation (\ref{eqT6}).
Two sets of expressions are appropriate to represent the \BiT,
depending on its later use.

The first set involves only the squares of the monodromy exponents,
\begin{eqnarray}
\frac{N}{u-U}
& = &
  \frac{x (x-1) U'}{U (U-1)(U-x)}
 +\left(\frac{\theta_0^2-\Theta_0^2}{N}-\frac{N}{4}  \right)\frac{1}{U}
 +\left(\frac{\theta_1^2-\Theta_1^2}{N}-\frac{N}{4}  \right)\frac{1}{U-1}
\label{eqTP6uveca}
\\
& &
\phantom{  \frac{x (x-1) U'}{U (U-1)(U-x)}}
 +\left(\frac{\theta_x^2-\Theta_x^2}{N}-\frac{N}{4}-1\right)\frac{1}{U-x}
\nonumber
\\
& = &
  \frac{x (x-1) u'}{u (u-1)(u-x)}
 +\left(\frac{\theta_0^2-\Theta_0^2}{N}+\frac{N}{4}  \right)\frac{1}{u}
 +\left(\frac{\theta_1^2-\Theta_1^2}{N}+\frac{N}{4}  \right)\frac{1}{u-1}
\\
& &
\phantom{  \frac{x (x-1) u'}{u (u-1)(u-x)}}
 +\left(\frac{\theta_x^2-\Theta_x^2}{N}+\frac{N}{4}-1\right)\frac{1}{u-x}\cdot
\nonumber
\end{eqnarray}

The second set of expressions, affine in the eight signed monodromy exponents,
\begin{eqnarray}
\frac{N}{u-U}
& = &
  \frac{x (x-1) U'}{U (U-1)(U-x)}
 +\frac{S_0\Theta_0}{U}+\frac{S_1\Theta_1}{U-1}+\frac{S_x\Theta_x-1}{U-x}
\\
& = &
  \frac{x(x-1)u'}{u(u-1)(u-x)}
 +\frac{s_0\theta_0}{u}+\frac{s_1\theta_1}{u-1}+\frac{s_x\theta_x-1}{u-x}
\ccomma
\label{eqTP6uvect}
\end{eqnarray}
is adapted to the construction of the contiguity relation, 
which is done in Section \ref{sectionDiscrete}.

Both sets allow one to directly apply the well known degeneracy
(Ref.~\cite{PaiCRAS1906}   in the space $\veca$,
 Ref.~\cite{Okamoto1986Pn} in the space $\vect$)
to generate {\BiT}s of the other $\Pn$ equations.
This is done in a forthcoming paper \cite{CM2001c}.

\section{Contiguity relation}
\label{sectionDiscrete}
\indent

{}From each \BiT,
one easily deduces a contiguity relation,
which generalizes, as noted by Garnier \cite{Garnier1951},
that of the hypergeometric equation of Gauss.
Its systematic computation is as follows \cite{FGR}.

\begin{enumerate}
\item
Consider the \BiT, i.e.~the direct \BiT\ and its inverse
\begin{eqnarray}
& &
u=f(U,U',x,\vect,\vecT),\
\vect=g(\vecT),
\label{eqSTDirect}
\\
& &
U=F(u,u',x,\vecT,\vect),\
\vecT=G(\vect).
\label{eqSTInverse}
\end{eqnarray}

\item
Evaluate it at the values 
$(\vminus,v,\vplus)$ taken by a discrete variable 
at three contiguous points $(z-h,z,z+h)$, with $z=n h$,
\begin{eqnarray}
& &
 \vplus=f(v,v',x,f(\vect),\vect),\
\\
& &
\vminus=F(v,v',x,F(\vect),\vect).\
\end{eqnarray}

\item
Eliminate the variable $v'$ between these two relations,
\begin{eqnarray}
& &
G(\vplus,\vminus,v,x,\vect)=0.
\end{eqnarray}

\end{enumerate}

For $\PVI$,
equations (\ref{eqSTDirect})--(\ref{eqSTInverse}) are equivalent to
\begin{eqnarray}
& &
\frac{x(x-1) U'}{U(U-1)(U-x)} =
 2 \frac{s_j \theta_j - S_j \Theta_j}{u-U}
-\left(
\frac{S_0\Theta_0}{U}+\frac{S_1\Theta_1}{U-1}+\frac{S_x\Theta_x-1}{U-x}\right),
\\
& &
\frac{x(x-1) u'}{u(u-1)(u-x)} =
-2 \frac{S_j \Theta_j - s_j \theta_j}{u-U}
-\left(
\frac{s_0\theta_0}{u}+\frac{s_1\theta_1}{u-1}+\frac{s_x\theta_x-1}{u-x}\right),
\end{eqnarray}
in which $j$ is anyone of the four singular points $(\infty,0,1,x)$,
and the contiguity relation is
\begin{eqnarray}
& &
\frac{\varphi(n+1/2)}{\vplus - v} + \frac{\varphi(n-1/2)}{\vminus - v}
=
 \frac{s_0 \theta_0 - S_0 \Theta_0}{v}
+\frac{s_1 \theta_1 - S_1 \Theta_1}{v-1}
+\frac{s_x \theta_x - S_x \Theta_x}{v-x}\ccomma
\label{eqContiguity}
\\
& &
\varphi(n)=
        \frac{1}{2}
(s_\infty \theta_\infty + s_0 \theta_0 + s_1 \theta_1 + s_x \theta_x -1),
\end{eqnarray}
in which $\vect$ is taken at the center point $z=z_0 + n h$.
This very simple expression is clearly invariant under any permutation of 
the four singular points of $\PVI$.

This contiguity relation (\ref{eqContiguity}) can be interpreted as a 
second order discrete equation \cite{FGR}.
The two-point recurrence relation (\ref{eqT6}) 
admits five classes of solutions.
Each class, characterized by a signature,
leads to a different contiguity relation (\ref{eqContiguity}),
i.e.~to a different second order discrete equation.
Four of them are autonomous 
(signatures $(s_j S_j)=$ $(---+)$, $(--++)$, $(-+++)$, $(++++)$),
they cannot admit a continuum limit to a Painlev\'e equation.
The only nonautonomous one (signature $(----)$)
is
\begin{eqnarray}
& &
\frac{n+1/2}{\vplus - v} + \frac{n-1/2}{\vminus - v}
=
 \frac{n + K_2 (-1)^n}{v}
+\frac{n + K_3 (-1)^n}{v-1}
+\frac{n + K_4 (-1)^n}{v-x}\ccomma
\label{eqContigP6a}
\\
& &
K_2=- k_2 + k_3 + k_4,\
K_3=  k_2 - k_3 + k_4,\
K_4=  k_2 + k_3 - k_4.
\label{eqContigP6b}
\end{eqnarray}
In the continuum limit,
among the six simple poles of $v$ in the sum (including $\infty$),
the first two will create a second order derivative 
and the four others will define at most four singular points.
Since none of the last four poles depends on $n$,
it is impossible that the continuum limit be $\PVI$.

The transform of this discrete equation under
\begin{eqnarray}
(\vplus,v,\vminus) \mapsto (\vplus,x/v,\vminus)
\end{eqnarray}
has already been obtained \cite[Eq.~(1.5)]{NRGO}    
as a reduction of a lattice KdV equation,
together with a discrete Lax pair 
and a continuum limit to the full $\PV$. 
This is in agreement with the continuum limit of the hypergeometric
contiguity relation,
which is not the hypergeometric equation but a confluent one.
Nevertheless,
we do not know of a general proof of this feature.

\section{On Schlesinger transformations}
\label{sectionOnST}

Up to this point we have carefully avoided using the expression 
\textit{Schlesinger transformation},
and this is because of some discrepancy which we would want
firstly, to point out,
secondly, to try to correct.

The relevant items are
\begin{enumerate}
\item
a birational transformation of some nonlinear ODE,
whose definition presents no ambiguity,

\item
the monodromy data,
also unambiguously defined by the isomonodromic deformation of some linear ODE,

\item
a Schlesinger transformation,
whose present definition, recalled below,
is in our opinion unsatisfactory.

\end{enumerate}

A \textit{Schlesinger transformation} (ST) was originally defined 
\cite{SchlesingerP6}
as a discrete transformation
preserving the monodromy of a given linear ODE,
and Schlesinger explicitly prescribed 
\cite[p.~136]{SchlesingerVorlesungen}
\cite[p.~134]{SchlesingerP6}
that monodromy exponents $\theta$
should be shifted by \textit{integer} values,
which is a sufficient condition to preserve the monodromy data.
All subsequent authors,
in particular Garnier \cite{Garnier1951} and 
Jimbo and Miwa \cite{JimboMiwaII}
(who created the expression Schlesinger transformation)
complied with this prescription
because all had in mind the isomonodromic deformations.

The discrepancy is the following.
It would be nice to have a one-to-one correspondence, hence an equivalence,
between the notion of birational transformation 
and that of Schlesinger transformation.
With the present definition,
this is not the case for the birational transformation (\ref{eqTP6uveca}),
whose associated affine transformation (\ref{eqT6})--(\ref{eqT9}) 
does not obey the prescription of Schlesinger of shifts by integers
and therefore does not preserve the monodromy data.
A first step in this direction was recently made 
in Refs.~\cite{DIKZ,KK1998,KMS},
who relaxed the prescription to shifts by \textit{half-integers},
so as to globally change the monodromy matrix to its opposite rather than to
conserve it.

What happens here is one step beyond:
the monodromy matrix is changed to its opposite,
but the shifts by half-integers occur not in the space of monodromy 
exponents,
but in the space adapted to the affine Weyl group ${\rm D}_4$ of $\PVI$.
Indeed,
in the basis of Okamoto (\ref{eqBasisbOkamoto}),
\begin{eqnarray}
& &
b_1=(\theta_0 + \theta_1)/2,\
b_2=(\theta_0 - \theta_1)/2,\
b_3=(\theta_x -1 - \theta_{\infty})/2,\
b_4=(\theta_x -1 + \theta_{\infty})/2,
\label{eqBasisbOkamoto}
\end{eqnarray}
the affine representation (\ref{eqT6}) becomes
(when all signs are chosen equal to $+1$ to simplify)
\begin{eqnarray}
& &
\TPVI : 
\pmatrix{b_1 \cr
         b_2 \cr
         b_3 \cr
         b_4 \cr}
= 
\pmatrix{-B_3 \cr
          B_2 \cr
         -B_1 \cr
          B_4 \cr}
\cdot
\label{eqT6basisb}
\end{eqnarray}
In the monodromy matrix,
the set of four trigonometric lines in Eq.~(3.19)
of Ref.~\cite{MS1995b} just undergoes a global sign change.

To conclude this discussion,
in order to identify Schlesinger transformation and birational transformation,
at least for $\PVI$ (we leave the full generality to mathematicians),
we propose that a \textit{Schlesinger transformation}
be defined as any discrete transformation which either conserves or changes
to its opposite the monodromy matrix,
without any additional prescription on the monodromy exponents.

As a consequence of this weaker definition,
the matrices $M_1$ and $M_0$ in (\ref{eqSTM1M0}) 
will have half-integer elements.
Another consequence of practical importance is
the identification, at least for the $\Pn$ equations,
of the two notions Schlesinger transformation, birational transformation,
since they will now have a one-to-one correspondence.

\section{Conclusion}
\indent

The fundamental homography between the derivative of the solution of the
Painlev\'e equation and the Riccati pseudopotential
has allowed us to define a consistent truncation.
The result thus obtained for $\PVI$ is a first degree \BiT,
and its contiguity relation defines
a single nonautonomous second order difference equation.
Its degeneracies under the confluence of the $\Pn$ equations
could provide new second order discrete equations.

As an application,
one can iterate this \BiT,
starting from
the two-parameter solution which Picard established for 
$\theta_j^{(0)}=0,j=\infty,0,1,x,$
to carry it to any $\vect$ such that $2 \theta_j$ 
and $\sum \theta_j$ be arbitrary integers.
Any iterate will be some algebraic transform of the solution of Picard.

Two important open problems remain in the domain of truncations,
namely to define truncations able to provide,
firstly  a Lax pair of $\PVI$, 
secondly a Lax pair 
for the discrete equations
built from the Schlesinger transformation.

\section*{Acknowledgments}
\indent

The authors acknowledge the financial support of the Tournesol grant
T99/040.
MM acknowledges the financial support of
the IUAP Contract No.~P4/08 funded by the Belgian government
and the support of CEA.

 \section {Appendix. One-family truncation for $\PVI$}
\label{appendixDetailedResultsOneFamilyTruncationPVI}

All families of movable singularities are equivalent under homographies.
The data for this unique representative family are
\begin{eqnarray}
& &
q=-6,\
u_0=-\frac{x(x-1)}{\theta_\infty},\
\hbox{ Fuchs index } 1,
\label{equ0}
\end{eqnarray}
so there are nine determining equations (\ref{eqST1Determining}) 
$E_j=0,j=0,\ldots,-q+2$.

The three coefficients $(A_2,A_1,A_0)$ are defined by (\ref{eqRiccatiUprime}),
\begin{eqnarray}
& &
A_2=
\frac{1}{2} \left[\frac{1}{U} + \frac{1}{U-1} + \frac{1}{U-x} \right],\
A_1=
- \left[\frac{1}{x} + \frac{1}{x-1} + \frac{1}{U-x} \right],\
\nonumber
\\
& &
A_0=
 \frac{U (U-1) (U-x)}{x^2 (x-1)^2} 
  \left[\Alpha + \Beta \frac{x}{U^2} + \Gamma \frac{x-1}{(U-1)^2} 
        + \Delta \frac{x (x-1)}{(U-x)^2} \right]\cdot
\end{eqnarray}

Equation $j=0$ only depends on $u_0$ and its solution is (\ref{equ0}).
Equation $j=1$ is identically satisfied,
as a consequence of the value $1$ of the Fuchs index.
There remain seven equations (\ref{eqST1Determining}) $j=2,\ldots,8$,
in the two unknowns $g_k(U,x),k=1,2$.
The next two ones are
\begin{eqnarray}
E_2 & \equiv &
 (g_2 \partial_U - \partial_x)g_1 +g_1^2 + g_0^{-1} F_0 g_1 + g_0^{-2} F_1=0,
\\
E_3 & \equiv &
 (-g_2^2 \partial_U^2 +2 g_2 \partial_U \partial_x - \partial_x^2) g_1
 + (-g_2^2 A_2 + g_2 A_1 - A_0 - 3 g_0) \partial_U g_1
\nonumber
\\
& &
 + 2 g_1^3 + g_0^{-1} F_2 g_1^2 + g_0^{-2} F_3 g_1 + g_0^{-3} F_4 =0,
\end{eqnarray}
in which the functions $F_k$ are differential polynomials of 
$(g_0,g_2,A_2,A_1,A_0)$,
\begin{eqnarray}
F_0 & \equiv &
\left(2 A_2 g_2 - A_1 + \theta_\infty \frac{2+2 x-6 U}{3 x(x-1)}\right) g_0
+ \partial_x g_0 - \partial_U (g_0 g_2),
\\
F_1 & \equiv &
48 \hbox{ terms}.
\end{eqnarray}

According to the conjecture of section \ref{sectionOneFamilyTruncation},
let us assume that $g_2$ is a second degree polynomial of $U$,
conveniently defined as
\begin{eqnarray}
& &
 g_2=U(U-1)(U-x)
\left(\frac{f_0(x)}{U} +\frac{f_1(x)}{U-1} + \frac{f_x(x)-1}{U-x}\right).
\label{eqg2Apriori}
\end{eqnarray}
To take full advantage of the symmetry of $\PVI$,
it is useful to also define
\begin{eqnarray}
& &
f_\infty=1-f_0-f_1-f_1.
\end{eqnarray}

Equation (\ref{eqg0Apriori}) then provides
\begin{eqnarray}
 g_0 & = &
  \frac{1}{2 (x(x-1))^2}
\Big\lbrace
 (\Theta_\infty^2 - f_\infty^2) U(U-1)(U-x)
-(\Theta_0^2 - f_0^2) \frac{x      (U-1)(U-x)}{U}
\nonumber \\ & &
\phantom{xxxxxxxxx}
+(\Theta_1^2 - f_1^2) \frac{ (x-1)U     (U-x)}{U-1}
-(\Theta_x^2 - f_x^2) \frac{x(x-1)U(U-1)     }{U-x}
\Big\rbrace
\nonumber \\ & &
+ \frac{1}{x(x-1)}\left(f_0' (U-1)(U-x) +f_1' U(U-x) +f_x' U(U-1)\right)\cdot
\label{eqg0PVI}
\end{eqnarray}

At this point,
one could enforce the second part of the conjecture,
i.e.~solve the three diophantine conditions on the rational functions $g_0$
and $g_1$ of $U$.
Let us instead proceed with the truncation.

Equation $j=2$ first provides the degrees in $U$ of the numerator and 
denominator of the rational function $g_1$.
Indeed, the degrees in $U$ of the polynomial coefficients of $E_2$ are
\begin{eqnarray}
\degree\left(U(U-1)(U-x)\right)^2 F_0=10,\
\degree\left(U(U-1)(U-x)\right)^4 F_1=20,\
\end{eqnarray}
therefore $U(U-1)(U-x)g_1$ is a fourth degree polynomial of $U$,
whose convenient definition implementing the invariance of $\PVI$ is
\begin{eqnarray}
& &
g_1= \Psi_\infty \frac{U}{x(x-1)} 
+ \frac{\Psi_0}{(x-1)U} - \frac{\Psi_1}{x(U-1)} + \frac{\Psi_x}{U-x} 
+ \frac{\Psi_c}{x(x-1)}
\ccomma
\label{eqg1PVI}
\end{eqnarray}
in which the five functions $\Psi_k$ only depend on $x$.
Equation $j=2$ then becomes a polynomial in $U$,
equivalently expanded in powers of $1/U$, $U$, $U-1$, or $U-x$,
\begin{eqnarray}
& &
E_2 
\equiv \sum_{k=0}^{20} E_{2,k}^{(\infty)} U^{20-k}
\equiv \sum_{k=0}^{20} E_{2,k}^{(0)} U^{k}
\equiv \sum_{k=0}^{20} E_{2,k}^{(1)} (U-1)^{k}
\equiv \sum_{k=0}^{20} E_{2,k}^{(x)} (U-x)^{k}=0,
\\
& &
\forall i \in \lbrace \infty,0,1,x \rbrace\ :\
E_{2,k}^{(i)}=0.
\end{eqnarray}
The four equations $k=0$ factorize,
\begin{eqnarray}
& &
j=2,\
k=0\ :
\left\lbrace
\matrix{
 \left(\Theta_\infty^2 - f_\infty^2\right)
&
 \left(\Theta_\infty^2-(f_\infty + 2 \Psi_\infty -2 \theta_\infty)^2\right)=0,
\hfill \cr
 \left(\Theta_0^2 - f_0^2\right)
&
 \left(\Theta_0^2 - (f_0 + 2 \Psi_0)^2\right)=0,
\hfill \cr
 \left(\Theta_1^2 - f_1^2\right)
&
 \left(\Theta_1^2 - (f_1 + 2 \Psi_1)^2\right)=0,
\hfill \cr
 \left(\Theta_x^2 - f_x^2\right)
&
 \left(\Theta_x^2 - (f_x + 2 \Psi_x)^2\right)=0,
\hfill \cr}
\right.
\end{eqnarray}
thus defining five possibilities since the signs of $\Theta_j$
are not prescribed.

The first possibility 
$\forall i \in \lbrace \infty,0,1,x \rbrace\ :\ \Theta_i^2-f_i^2=0$
is ruled out since the $\Theta_i$'s must remain arbitrary.

The second possibility,
that one $\Theta_i^2-f_i^2$ be nonzero and the three others vanish,
defines four equivalent subcases, e.g.
\begin{eqnarray}
& &
f_0=\Theta_0,\
f_1=\Theta_1,\
f_x=\Theta_x,\
\Psi_\infty=\theta_\infty- \frac{1}{2}
       \left(\Theta_\infty - \Theta_0 - \Theta_1 - \Theta_x + 1\right).
\label{eqj2k0}
\end{eqnarray}
At $j=2,k=4$, 
one obtains
\begin{eqnarray}
& &
{\hskip -14.0truemm}
j=2,\ k=4\ :
\Psi_c=-\frac{x+1}{3} \Psi_\infty,\
\Psi_0 (\Psi_0 - \Theta_0)=0,\
\Psi_1 (\Psi_1 - \Theta_1)=0,\
\Psi_x (\Psi_x - \Theta_x)=0.
\end{eqnarray}
At $j=2,k=5$, 
the subcase $(\Psi_0 - \Theta_0)(\Psi_1 - \Theta_1)(\Psi_x - \Theta_x)=0$
is ruled out and one obtains
\begin{eqnarray}
& &
12 \theta_0^2 - 4 \theta_\infty^2 =
 3 \left(-\Theta_\infty + \Theta_0 - \Theta_1 - \Theta_x + 1\right)^2
 - \left( \Theta_\infty - \Theta_0 - \Theta_1 - \Theta_x + 1\right)^2,\
\\
& &
12 \theta_1^2 - 4 \theta_\infty^2 =
 3 \left(-\Theta_\infty - \Theta_0 + \Theta_1 - \Theta_x + 1\right)^2
 - \left( \Theta_\infty - \Theta_0 - \Theta_1 - \Theta_x + 1\right)^2,\
\\
& &
12 \theta_x^2 - 4 \theta_\infty^2 =
 3 \left(-\Theta_\infty - \Theta_0 - \Theta_1 + \Theta_x + 1\right)^2
 - \left( \Theta_\infty - \Theta_0 - \Theta_1 - \Theta_x + 1\right)^2.
\end{eqnarray}
This exhausts the equation $j=2$.
Then, the equation $j=3$ is a polynomial of degree $3$ in $U$,
and the equation $E_{3,k}^{(0)}=0$ provides the last information,
\begin{eqnarray}
& &
\theta_\infty = \frac{1}{2}
  \left( \Theta_\infty - \Theta_0 - \Theta_1 - \Theta_x + 1\right).
\end{eqnarray}
Therefore, this second possibility
(one $\Theta_i^2-f_i^2$ nonzero and the three others equal to zero)
defines the solution
\begin{eqnarray}
& &
{\hskip -12.0truemm}
g_0=\frac{N U (U-1)(U-x)}{u_0 x(x-1)},\
g_1=0,\
g_2=
\left(\frac{\Theta_0}{U}+\frac{\Theta_1}{U-1}+\frac{\Theta_x-1}{U-x}\right)
\frac{g_0 u_0}{N},\
u_0=-\frac{x(x-1)}{\theta_\infty}\cdot
\end{eqnarray}
With the three choices other than (\ref{eqj2k0}),
one would simply obtain the three solutions deduced from that one
by applying to the \RHS\ of (\ref{eqSTDirecte})
the homographies $\Hbadc,\Hdcba,\Hcdab$,
solutions characterized by the following values of $g_1$,
\begin{eqnarray}
& &
u_0 g_1=
0,\
\frac{x}{U}-U,\
x + \frac{x(x-1)}{U-x}-U,\
1 + \frac{1-x}   {U-1}-U.
\end{eqnarray}

The third, fourth and fifth possibilities
of vanishing of the four $\Theta_i^2-f_i^2$ are currently under examination,
just to be sure that they provide no other solution.

The full results are given in the text.



\vfill \eject

\end{document}